**Manuscript**

# Enhanced Phonon-Phonon Interactions and Weakened Electron-Phonon Coupling in Charge-Density-Wave Topological Semimetal $EuAl_4$ with a Possible Intermediate Electronic State

*Shize Cao,*[1,2] *Feng Jin,*[1] *Jianzhou Zhao,*[3] *Yun-Ze Long,*[2,*] *Jianlin Luo,*[1,4] *Qingming Zhang*[1,5]

*and Zhi-Guo Chen*[1,4,*]

[1] Beijing National Laboratory for Condensed Matter Physics, Institute of Physics, Chinese Academy of Sciences, Beijing 100190, China

[2] Collaborative Innovation Center for Nanomaterials & Devices, College of Physics, Qingdao University, Qingdao 266071, China

[3] Co-Innovation Center for New Energetic Materials, Southwest University of Science and Technology, Mianyang 621010, China

[4] Songshan Lake Materials Laboratory, Dongguan, Guangdong, 523808, China

[5] School of Physical Science and Technology, Lanzhou University, Lanzhou 730000, China

ABSTRACT: The origin of charge density wave (CDW) is a long-term open issue. Furthermore, the evolution of phonon-phonon interactions (PPI) across CDW transitions has rarely been investigated. Besides, whether electron-phonon coupling (EPC) would be weakened or enhanced after CDW transitions is still under debate. Additionally, CDW provides a fertile ground for uncovering intriguing intermediate electronic states. Here, we report a Raman spectroscopy study of the PPI and EPC in topological semimetal $EuAl_4$ exhibiting a CDW phase below temperature $T_c$ ~ 145 K. The free-charge-carrier-density ($n_c$) and temperature dependences of the Fano asymmetric factors ($1/|q|$) of the two phonon modes $A_{1g}$ and $B_{1g}$ indicates that below $T_c$, the EPC becomes weakened probably due to the reduction of the $n_c$. Interestingly, in the temperature range 50—145 K, the steep growth of the $1/|q|$ leading to the significant deviation from the linear dependence on the $n_c$, together with the shoulder-like features in the temperature evolutions of the $1/|q|$ and the $n_c$ around 50 K, implies the possible existence of an intermediate electronic state with the EPC distinctly larger than the CDW ground state in $EuAl_4$. Furthermore, below $T_c$, the faster decrease in the full width at half maxima of the $B_{1g}$ phonon mode representing the collective vibrations of the CDW-modulated $Al^1$ atoms suggests that a remarkable growth of the PPI for the $B_{1g}$ phonon mode after the CDW phase transition, which is in contrast to the weakening of the EPC and thus may mainly arise from the strengthening of lattice anharmonicity in $EuAl_4$. Our results not only highlight the significance of the enhanced PPI and the weakened EPC in completely understanding the formation of the CDW phase but also initiate the exploration of novel intermediate electronic states in $EuAl_4$.

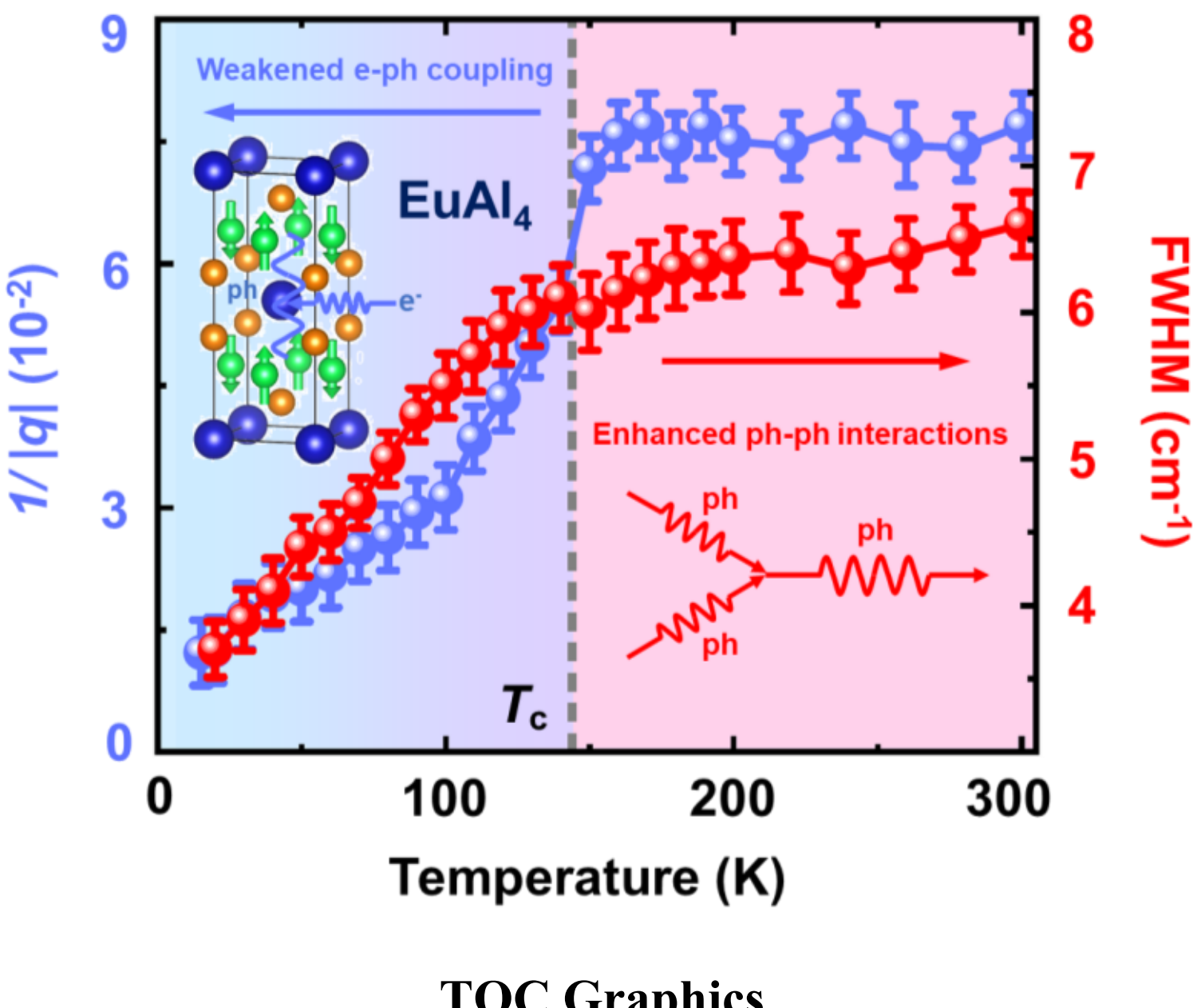

**TOC Graphics**

Charge density wave (CDW), which represents a quantum state of matter with the charge density and underlying lattice periodically modulated, has generated enormous interest in the scientific community because it provides a platform for realizing a variety of quantum phenomena, such as electronic nematic state[1-4], axion insulator state[5-6], superconductivity[7-11] and anomalous Hall effect[12]. Therein, electronic nematic state, which was revealed to show an intriguing entanglement with unconventional superconductivity[13-18], is a type of intermediate electronic states adjacent to CDW ground state. Revealing the mechanism of the formation of CDW states is an important step towards not only searching for new CDW materials but also achieving interesting quantum phenomena. However, the driving mechanism for the formation of CDW remains elusive. For one-dimensional systems, the Fermi surface nesting was proposed to cause the divergence of the Lindhard response function, which can result in the presence of CDW order[19-20]. Nonetheless, since small deviation from perfect

Fermi nesting would greatly suppress the divergence of the Lindhard response function, Fermi-surface-nesting induced CDW were observed in few realistic materials. Many-body interactions including phonon-phonon interactions (PPI) and electron-phonon coupling (EPC) play crucial roles in the emergence of various quantum states in condensed matter[21-22]. Therein, a sufficiently strong momentum-dependent EPC was suggested to trigger Kohn anomaly (or phonon softening), which can also induce a CDW phase transition[23]. It is a pity that a consensus on the evolution of the strength of EPC across CDW transitions has not been reached, e.g., the EPC strength decreases in $ZrTe_3$[24] but increases in $La_{1.8-x}Eu_{0.2}Sr_xCuO_{4+\delta}$ [22] upon entering the CDW state. In addition, the change in the strength of phonon-phonon interactions (PPI) across CDW transitions has rarely been reported[21].

A rare-earth intermetallic $EuAl_4$ crystallizes in the centrosymmetric tetragonal structure (the space group *I*4/*mmm*) with two types of Al atoms—$Al^1$ and $Al^2$. Below a critical temperature $T_c$ ~ 145 K, $EuAl_4$ was revealed to undergo an incommensurate CDW order with the positions of the $Al^1$ and Eu atoms modulated[25-30]. Besides, at $T_\chi$ ~ 10.1 K in the helical spin-density-wave state of $EuAl_4$, a first-order phase transition gives rise to a spontaneous flipping of spin chirality[31-32]. Moreover, nontrivial topology including topological non-trivial bands in momentum space[33-34] and topological spin textures—skyrmion in real space was proposed to exist in $EuAl_4$[35]. Furthermore, topological Hall effect is likely to accompany with the CDW in $EuAl_4$[36]. Therefore, $EuAl_4$ can serve as a platform for studying the interplay between CDW and nontrivial topology. Clarifying the microscopic mechanism for the formation of the CDW entangled with nontrivial topology in $EuAl_4$ would lay a solid foundation for exploring exotic effects arising from the combination of CDW and nontrivial topology. Previous investigations indicate that the softening of a transverse acoustic phonon mode in $EuAl_4$ is a crucial driving force for the presence of its CDW order[37]. Nonetheless, the evolutions of the EPC and PPI across CDW transitions in $EuAl_4$, which is expected to offer significant information for completely understanding microscopic mechanism for its CDW formation, were seldom investigated by experiments.

Raman spectroscopy is an efficient experimental technique for probing phonon modes which include the information about PPI and EPC in solids[21, 24, 38-41]. Thus, a Raman spectroscopy study of the temperature evolutions of the Raman-active phonons across the CDW transition in topological semimetal $EuAl_4$, which has yet been performed up to now, would shed light on the evolutions of PPI and EPC. Here, we performed the Raman spectroscopy measurements of the (001) faces of the $EuAl_4$ single crystals in the temperature range from 15 to 300 K in the backscattering configuration on a Horiba Jobin Yvon LabRAM HR Evolution micro-Raman spectrometer using the 473 nm laser as the light source (see the details about the single-crystal growth, and the X-ray-diffraction characterization in Section 1 of Supporting Information). The $EuAl_4$ single crystals were cooled using low-temperature helium exchange gas in an AttoDRY 2100 cryostat.

**Figure 1a** and **1b** depict two asymmetric peak-like features around 247 $cm^{-1}$ and 390 $cm^{-1}$ in the Raman spectra of the $EuAl_4$ single crystals at different temperatures. To study the nature of the observed peak-like features, we analyzed the *I*4/*mmm* space group of its crystal structure.

The phonon modes of $EuAl_4$ at the Γ point can be decomposed into the following irreducible representations: $[A_{1g} + B_{1g} + 2E_g] + [2A_{2u} + 2E_u] + [A_{2u} + E_u]$, where the first, second, and third terms represent the Raman-active, infrared-active, and the acoustic phonon modes, respectively. The corresponding Raman tensors of the *I*4/*mmm* space group can be expressed as:

$$A_{1g} = \begin{pmatrix} a & 0 & 0 \\ 0 & a & 0 \\ 0 & 0 & b \end{pmatrix} \qquad B_{1g} = \begin{pmatrix} c & 0 & 0 \\ 0 & -c & 0 \\ 0 & 0 & 0 \end{pmatrix} \tag{1}$$

$$E_g = \begin{pmatrix} 0 & 0 & 0 \\ 0 & 0 & d \\ 0 & d & 0 \end{pmatrix} \qquad E_g = \begin{pmatrix} 0 & 0 & -d \\ 0 & 0 & 0 \\ -d & 0 & 0 \end{pmatrix} \tag{2}$$

where the *a*, *b*, *c* and *d* are the tensor elements and their values depend on the cross-section of Raman scattering. We further performed *ab initio* calculations of the phonon spectrum of $EuAl_4$ (see **Figure 2a** and the details about our *ab initio* calculations in Section 1 of Supporting Information). The yellow and green dots in **Figure 2a** indicate that the two phonon bands at the Γ point, which separately host the symmetry $B_{1g}$ and $A_{1g}$, exhibit the energies: 232.4 $cm^{-1}$ and 399.4 $cm^{-1}$. Considering that the energies of the two peak-like features in the measured Raman spectra of the $EuAl_4$ single crystals are quite close to (i) those of the calculated phonon modes $B_{1g}$ around 232.4 $cm^{-1}$ and $A_{1g}$ around 399.4 $cm^{-1}$ of $EuAl_4$ (see **Table 1**), and (ii) those of the previously reported phonon modes $B_{1g}$ (around 229.7 $cm^{-1}$) and $A_{1g}$ (around 359.9 $cm^{-1}$) of its counterpart $SrAl_4$ with isostructure and isovalence[34], the asymmetric peak-like features around 247 $cm^{-1}$ and 390 $cm^{-1}$ in **Figure 1a** and **1b** can be assigned as the $B_{1g}$ and $A_{1g}$ phonon modes, respectively. Therein, the $B_{1g}$ and $A_{1g}$ phonon modes separately represent the collective vibrations of the $Al^1$ and $Al^2$ atoms (see **Figure 2b** and **2c**).

As is well known, fitting an asymmetric phonon mode using Breit-Wigner-Fano (i.e., BWF) line shape can yield the asymmetric factor $1/|q|$, which quantifies the strength of EPC. To obtain the temperature dependences of the EPC strength of the asymmetric $B_{1g}$ and $A_{1g}$ phonon modes, we fit these two phonon modes using BWF line shapes[42-43]:

$$I(\omega) = I_0 \frac{\left(1 + \frac{\omega - \omega_{ph}}{q\Gamma}\right)^2}{1 + \left(\frac{\omega - \omega_{ph}}{\Gamma}\right)^2} \tag{3}$$

where $I_0$, $\omega_{\mathrm{ph}}$, $\Gamma$ and $1/|q|$ are the maximal intensity, energy, full width at half maxima (i.e., FWHM), and asymmetric factors of the BWF line shapes, respectively (see the fitting parameters at different temperatures in **Table S1** of Supporting Information). The upper and lower panels of **Figure 3a** and **3b** depict the BWF fits to the $B_{1g}$ and $A_{1g}$ phonon modes at two typical temperatures $T$ = 15 K and 300 K, respectively. The asymmetric factors (i.e., $1/|q|(B_{1g}$, 15 K) ~ 0.013 and $1/|q|(A_{1g}$, 15 K) ~ 0.012) for the BWF fits to the $B_{1g}$ and $A_{1g}$ phonon modes at $T$ = 15 K is distinctly smaller than those (i.e., $1/|q|(B_{1g}$, 300 K) ~ 0.075 and $1/|q|(A_{1g}$, 300 K) ~ 0.055) at $T$ = 300 K, which suggests that the EPC in $EuAl_4$ at $T$ = 15 K is weaker than that at $T$ = 300 K. To get the temperature evolution of the EPC, we plotted the two $1/|q|$ for the $B_{1g}$ and $A_{1g}$ phonon modes in the temperature range from 15 to 300 K in **Figure 3c**. Above $T_c$, the two $1/|q|$ show weak temperature dependences, but below $T_c$, the two $1/|q|$ decrease sharply with lowering temperature, which indicates that EPC is weakened after the CDW transition.

**Figure 3c** displays that the previously reported relative Drude weight $S_D(T)/S_D(300\ \mathrm{K})$ of $EuAl_4$ which is proportional to the free charge carrier density $n_c$ as a function of temperature[44]. The significant suppression of the $S_D(T)/S_D(300\ \mathrm{K})$ below $T_c$ manifests that the free charge carrier density is dramatically lowered after the CDW phase transition in $EuAl_4$. **Figure 3d** shows that the two $1/|q|$ of the $B_{1g}$ and $A_{1g}$ phonon modes decrease with the lowering of the free charge carrier density, which suggests that the weakening of EPC after the CDW transition in $EuAl_4$ is likely to be caused by the reduction of the carrier density. In addition, as shown by the dashed line in **Figure 3d**, both the two $1/|q|$ of the $B_{1g}$ and $A_{1g}$ phonon modes below 50 K can be linearly extrapolated to zero at $S_D(T)/S_D(300\ K) = 0$ (see the black dashed line in **Figure 3d**), which implies that in $EuAl_4$, the vanishing of free charge carriers would make the $B_{1g}$ and $A_{1g}$ phonon modes recover their Lorentzian line shapes with $1/|q| = 0$. Surprisingly, in the temperature from 50 to 145 K, on the right side of the red dashed vertical line in **Figure 3d**, the two $1/|q|$ of the $B_{1g}$ and $A_{1g}$ phonon modes grow steeply and deviate significantly from the linear dependence on the $S_D(T)/S_D(300\ K)$. It is well known that the BWF line shapes of the phonon modes in solids arise from the coupling between phonons and electron-hole pair excitations due to EPC[42-43]. The abrupt growth of the two $1/|q|$ of the $B_{1g}$ and $A_{1g}$ phonon modes with increasing temperature from 50 to 145 K suggests that in $EuAl_4$, an intermediate electronic state with distinctly larger EPC than the incommensurate CDW ground state may exist in the temperature range 50—145 K, which is supported by the presence of the shoulder-like features around $T \sim 50$ K in the temperature dependences of the $1/|q|$ here (see the purple down arrow in **Figure 3c**). Moreover, if there is no intermediate electronic state in a CDW material, its Drude weight would drop quickly and continuously with the absence of a shoulder-like feature in the temperature dependence of the Drude weight after the CDW phase transition [45]. However, a shoulder-like feature is present around $T \sim 50$ K in the temperature dependences of the previously reported $S_D(T)/S_D(300\ \mathrm{K})$ (see the red up arrow in **Figure 3c**), which further supports the possible intermediate electronic state in $EuAl_4$.

To estimate the evolution of the PPI across the CDW phase transition in $EuAl_4$, we studied the temperature evolutions of the FWHMs and energies of its $B_{1g}$ and $A_{1g}$ phonon modes. **Figure 4a** and **4b** show that the FWHM of the $B_{1g}$ phonon mode drops more fastly below $T_c$ than above $T_c$, while the FWHM of the $A_{1g}$ phonon mode shows unobvious anomaly. Given the anharmonic and harmonic contributions to the FWHMs of the $B_{1g}$ and $A_{1g}$ phonons, the temperature dependence of the FWHM of $A_{1g}$ phonon above and below $T_c$ can be fit well using one PPI parameter ($\lambda_{\text{ph-ph}} \sim 0.38$) in the temperature range 15—300 K based on the following relationship[21, 46-47]:

$$\Gamma(T) = \Gamma_0 \left[ 1 + \frac{2\lambda_{ph-ph}}{\exp\left(\frac{\hbar\omega_0}{2k_B T}\right) - 1} \right] \tag{4}$$

where $\omega_0$ and $\Gamma_0$ are the energy and linewidth of phonon at 0 K (see the fitting parameters in **Table** S2 of Supporting Information). By contrast, the PPI parameter for fitting the temperature evolution of the $B_{1g}$ phonon FWHM increases from 0.04 to 0.31 upon entering the CDW state (see the fitting parameters in **Table** S2). Considering that the $B_{1g}$ phonon, rather than the $A_{1g}$ phonon, corresponds to the collective vibrations of the $Al^1$ atoms which are modulated periodically after the CDW transition, the CDW phase transition in $EuAl_4$ leads to a remarkable growth of the PPI for the $B_{1g}$ phonon below $T_c$.

To check the enhancement of the phonon-phonon interactions in $EuAl_4$ after the CDW transition, we plotted the energies of the $B_{1g}$ and $A_{1g}$ phonon modes as a function of temperature in **Figure 4c** and **4d**, respectively. Generally, phonon-mode energy can be given by the following relationships which contain anharmonic ($\omega_{ah}(T)$), quasi-harmonic ($\omega_{qh}(T)$) and harmonic (i.e., $\omega_h$) terms[21, 46-47]:

$$\omega(T) = \omega_h + \omega_{qh}(T) + \omega_{ah}(T) \tag{5}$$

$$\omega_{ah}(T) = -\omega_0 \left\{ \left( \frac{\Gamma_0}{\sqrt{2}\omega_0} \right)^2 \left[ 1 + \frac{4\lambda_{ph-ph}}{\exp\left( \frac{\hbar\omega_0}{2k_B T} \right) - 1} \right] \right\} \tag{6}$$

$$\omega_{qh}(T) = \omega_0 \left\{ \exp\left[ -3\gamma \int_0^T \alpha_i(T') dT' \right] - 1 \right\} \tag{7}$$

where $\gamma$ is the Grüneisen parameter of a phonon mode and $\alpha_i(T')$ is the coefficient of the linear thermal expansion at temperature $T'$. Here, the temperature dependences of the quasi-harmonic terms of the $B_{1g}$ and $A_{1g}$ phonon energies are given by the product of the Grüneisen constant $\gamma$ and the integral of the linear expansion coefficient up to a temperature $T$ (see the integral of the linear expansion coefficient in Section 2 and **Figure S1** of Supporting Information)[48]. Therein, the temperature dependence of the linear expansion coefficient of $EuAl_4$ were obtained by the previously reported thermodynamic investigation. Thus, fitting the temperature dependences of the $B_{1g}$ and $A_{1g}$ phonon energies based on Equation (5)-(7) yields the Grüneisen constants (for the $B_{1g}$ phonon, $\gamma_{B_{1g}}$ = 1.9 above $T_c$ and $\gamma_{B_{1g}}$ = 6.1 below $T_c$ and for the $A_{1g}$ phonon, $\gamma_{A_{1g}}$ = 1.8 above and below $T_c$) and the PPI parameters (for the $B_{1g}$ phonon, $\lambda_{ph-ph}^{B_{1g}}$= 0.04 above $T_c$ and $\lambda_{ph-ph}^{B_{1g}}$ = 0.31 below $T_c$, while for the $A_{1g}$ phonon, $\lambda_{ph-ph}^{A_{1g}}$= 0.38 above and below $T_c$), which are consistent with the PPI parameters obtained via fitting the temperature dependences of the FWHMs of the $B_{1g}$ and $A_{1g}$ phonons based on Equation (4) (see the fitting parameters in **Table** S2 of Supporting Information).

It is worth noticing that PPI not only can be mediated by electrons due to EPC[49] but also can come from lattice anharmonicity[21, 47, 50]. If the PPI in $EuAl_4$ are mediated by electrons, the decrease in the EPC strengths estimated from the two $1/|q|$ of the $B_{1g}$ and $A_{1g}$ phonon modes upon entering the CDW state should correspond to the weakening of the PPI. On the contrary, after the CDW phase transition in $EuAl_4$, the PPI parameter for the $B_{1g}$ phonon is enhanced from 0.04 to 0.31. Thus, the PPI in $EuAl_4$ should mainly arise from lattice anharmonicity. Moreover, the remarkable growth of the PPI for the $B_{1g}$ phonon below $T_c$ implies the strengthening of lattice anharmonicity after the CDW phase transition in $EuAl_4$.

In summary, using Raman spectroscopy, we have investigated the PPI and EPC in CDW topological semimetal $EuAl_4$. Two asymmetric peak-like features in the Raman spectra of the $EuAl_4$ single crystals are assigned to the Raman-active phonon modes $A_{1g}$ and $B_{1g}$. Below $T_c$, the reduction of the Fano asymmetric factors ($1/|q|$) of the two phonon modes $A_{1g}$ and $B_{1g}$ indicates that the EPC becomes weakened after the CDW phase transition. Moreover, the decrease in the Fano asymmetric factors ($1/|q|$) is accompanied with the suppression of free charge carrier density ($n_c$), which suggests that the weakening of the EPC in $EuAl_4$ is likely caused by the reduction of the $n_c$. Unexpectedly, in the temperature range 50—145 K, the two $1/|q|$ grow steeply and deviate significantly from the linear dependences on the $n_c$. The steep growth of the $1/|q|$, combined with the shoulder-like features in the temperature evolutions of the $1/|q|$ and the $n_c$ around 50 K, implies the possible existence of an intermediate electronic state with the EPC distinctly larger than the CDW ground state in $EuAl_4$. Furthermore, the FWHM of the $B_{1g}$ phonon mode representing the collective vibrations of the CDW-modulated $Al^1$ atoms decreases more fastly below $T_c$ than above $T_c$, which manifests that a remarkable growth of the PPI for the $B_{1g}$ phonon mode after the CDW phase transition. Since the remarkable growth of the PPI for the $B_{1g}$ phonon mode is in contrast to the weakening of the EPC below $T_c$, the PPI enhancement may mainly originate from the strengthening of lattice anharmonicity in $EuAl_4$. Our work not only establishes the importance of the enhanced PPI and the weakened EPC for revealing the formation mechanism of the CDW phase but also offers a clue to discovering exotic intermediate electronic states in $EuAl_4$.

## Supporting Information

Single crystals growth, structural characteristics, fitting details about the quasiharmonic and anharmonic models, details about the *ab initio* calculations of the phonon spectrum and fitting parameters for the BWF, quasiharmonic and anharmonic models.

## Author Information

### Corresponding Authors

**Zhi-Guo Chen** – Beijing National Laboratory for Condensed Matter Physics, Institute of Physics, Chinese Academy of Sciences, Beijing 100190, China; Email: zgchen@iphy.ac.cn

**Yun-Ze Long** – Collaborative Innovation Center for Nanomaterials & Devices, College of Physics, Qingdao University, Qingdao 266071, China; Email: yunze.long@qdu.edu.cn

**Authors**

**Shize Cao** – Beijing National Laboratory for Condensed Matter Physics, Institute of Physics, Chinese Academy of Sciences, Beijing 100190, China

**Feng Jin** – Beijing National Laboratory for Condensed Matter Physics, Institute of Physics, Chinese Academy of Sciences, Beijing 100190, China

**Jianzhou Zhao** – Co-Innovation Center for New Energetic Materials, Southwest University of Science and Technology, Mianyang 621010, China

**Jianlin Luo** – Beijing National Laboratory for Condensed Matter Physics, Institute of Physics, Chinese Academy of Sciences, Beijing 100190, China

**Qingming Zhang** – Beijing National Laboratory for Condensed Matter Physics, Institute of Physics, Chinese Academy of Sciences, Beijing 100190, China

**Author Contributions**

Z.-G.C. conceived and supervised this project; S.C. grew the single crystals; F.J. and Q.Z. built up the Raman measurement system; J.Z. performed the *ab initio* calculations of the phonon spectrum; S.C. carried out the Raman measurements with the assistance of F.J.; Z.-G.C., S.C., Y.L. and J.L. analyzed the data; Z.-G.C. wrote the paper with the input from S.C..

**Notes**

The authors declare no conflict of interest.

## Acknowledgements

The authors acknowledge support from the Guangdong Basic and Applied Basic Research Foundation (Project No. 2021B1515130007), the National Natural Science Foundation of China (Project No. U21A20432 and 12274186), the strategic Priority Research Program of Chinese Academy of Sciences (Project No. XDB33000000), the National Key Research and Development Program of China (Grant No. 2022YFA1403800, 2024YFA1408301 and 2022YFA1402704) and the Synergetic Extreme Condition User Facility (SECUF, https://cstr.cn/31123.02.SECUF).

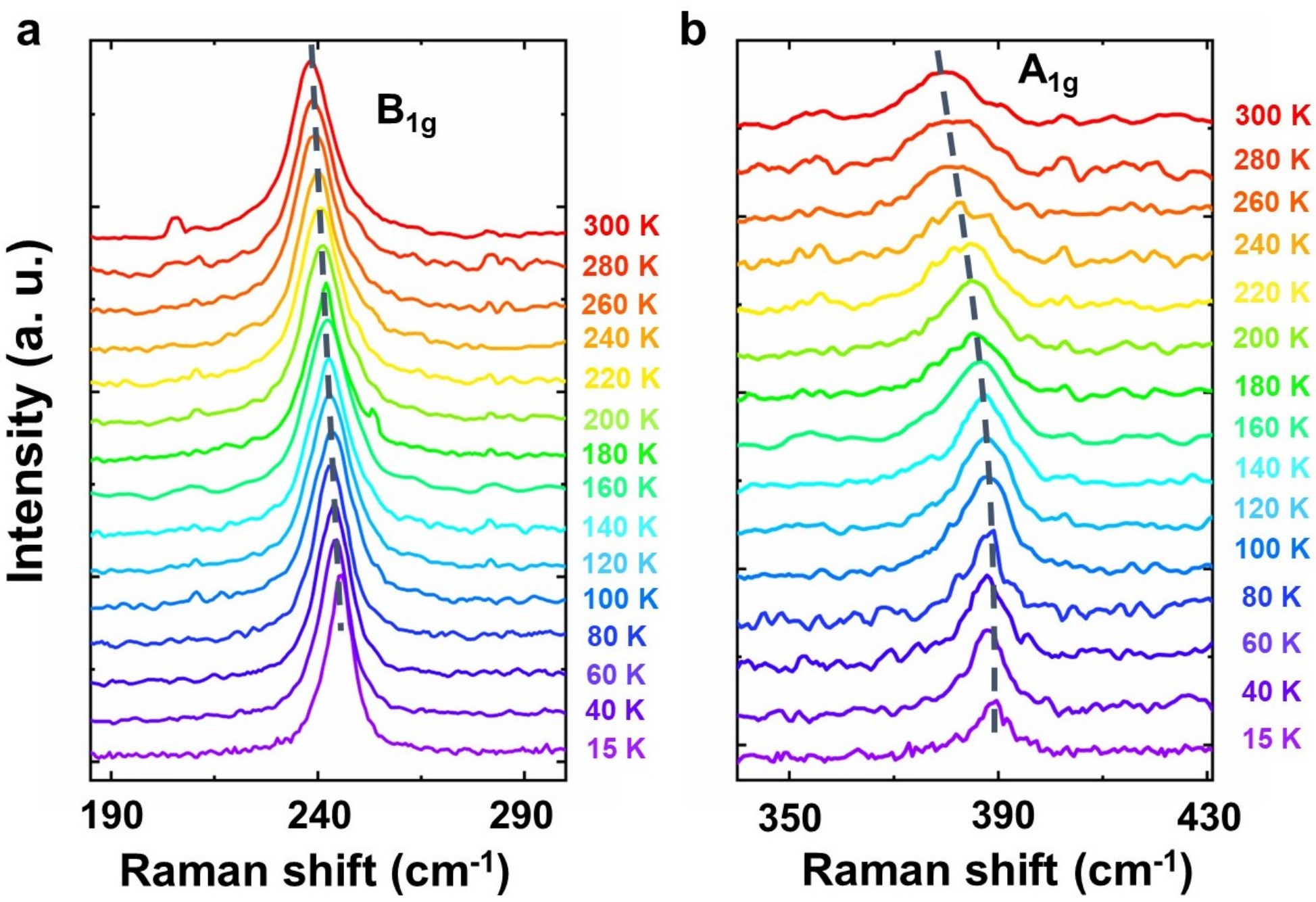

**Figure 1.** Two asymmetric Raman-active phonon modes $B_{1g}$ and $A_{1g}$ in the Raman spectra of $EuAl_4$ at different temperatures. The dashed curves in **a**) and **b**) are guides to the eye for the temperature evolutions of the $B_{1g}$ and $A_{1g}$ phonon modes, respectively.

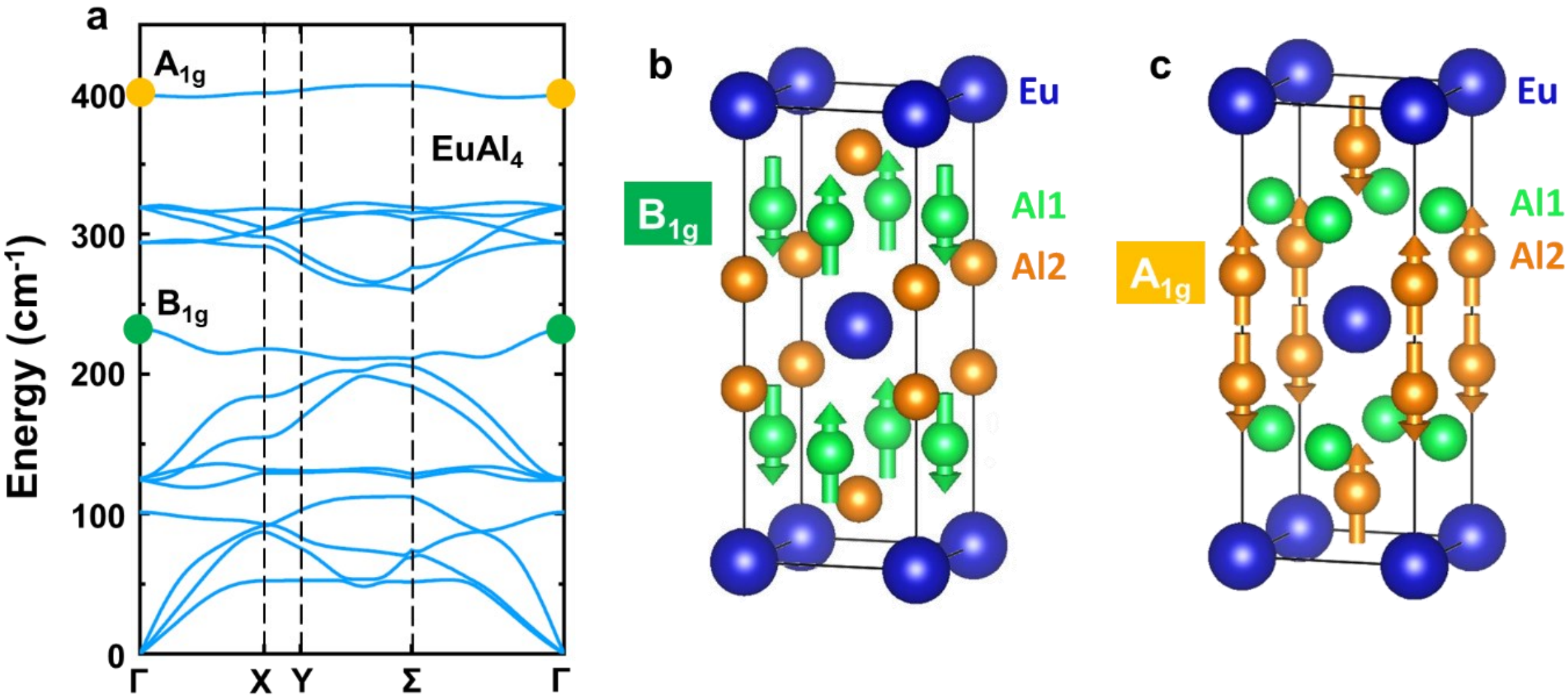

**Figure 2.** Theoretically calculated Raman-active phonon modes $B_{1g}$ and $A_{1g}$. **a)** Phonon spectrum of $EuAl_4$ obtained by *ab initio* calculations. The yellow and green dots in a) indicate the calculated energies of the Raman-active phonon modes $A_{1g}$ and $B_{1g}$. **b)** and **c)** Schematics of the collective vibrations of the $Al^1$ and $Al^2$ atoms in $EuAl_4$, which separately correspond to the $B_{1g}$ and $A_{1g}$ phonon modes.

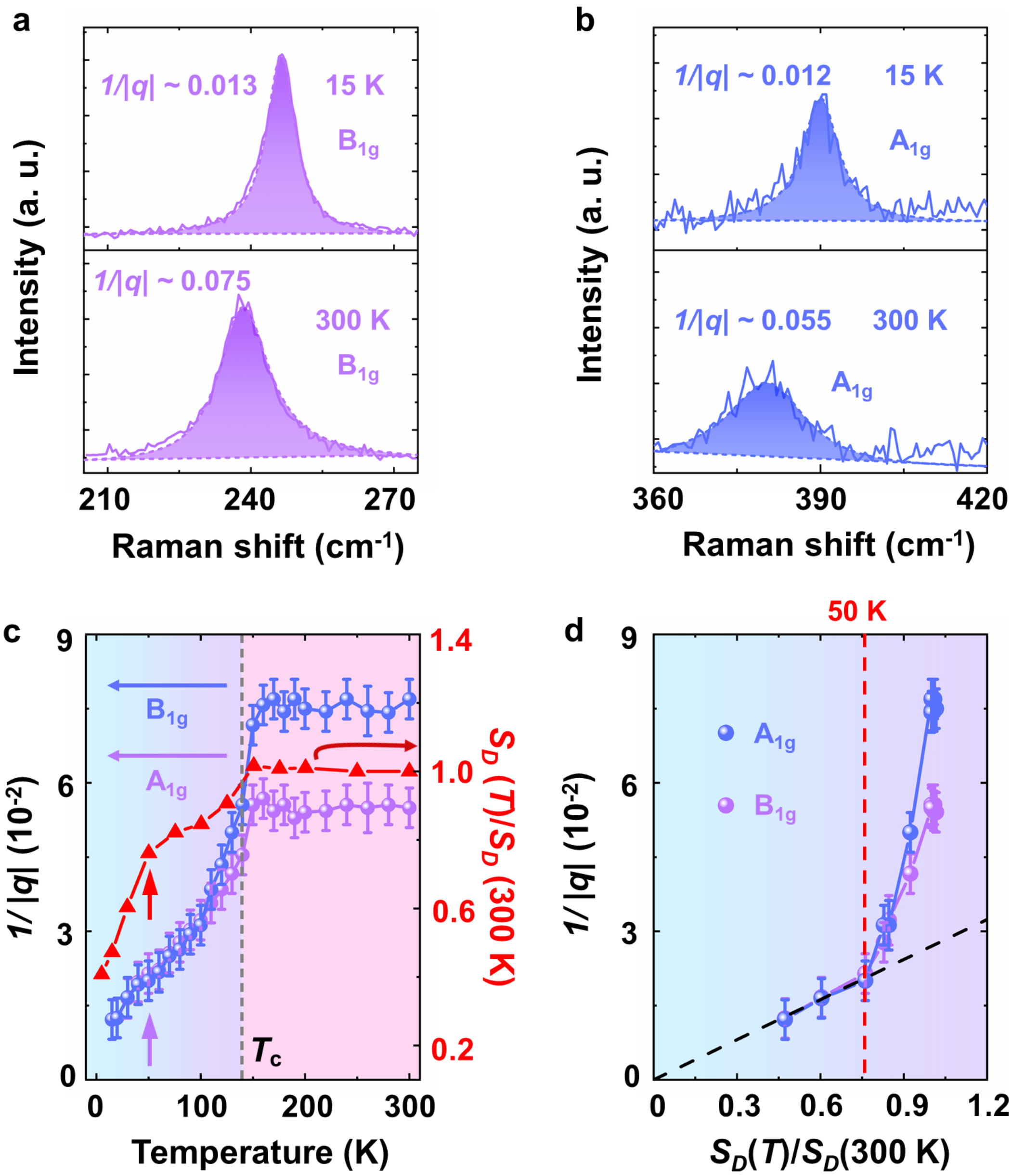

**Figure 3.** Strength of the Electron-phonon coupling in $EuAl_4$. **a**) BWF fits to the $B_{1g}$ phonon mode at $T$ = 15 K and 300 K. **b**) BWF fits to the $A_{1g}$ phonon mode at $T$ = 15 K and 300 K. c) Temperature dependences of the asymmetric factors $1/|q|$ for the $B_{1g}$ and $A_{1g}$ phonon modes and the previously reported relative Drude weight $S_D(T)/S_D$(300 K). The purple down arrow and the red up arrow in **c**) indicate the shoulder-like features in the temperature dependences of the $1/|q|$ and the $S_D(T)/S_D$(300 K) around 50 K, respectively. **d**) Two $1/|q|$ of the $B_{1g}$ and $A_{1g}$ phonon modes as a function of the $S_D(T)/S_D$(300 K). The red dashed vertical line in d) shows that on its right side (i.e., in the temperature range 50—145 K), the $1/|q|$ of the $B_{1g}$ and $A_{1g}$ phonon modes grow steeply and deviate significantly from the linear dependence on the $S_D(T)/S_D$(300 K) (see the black dashed line), which can be linearly extrapolated to zero.

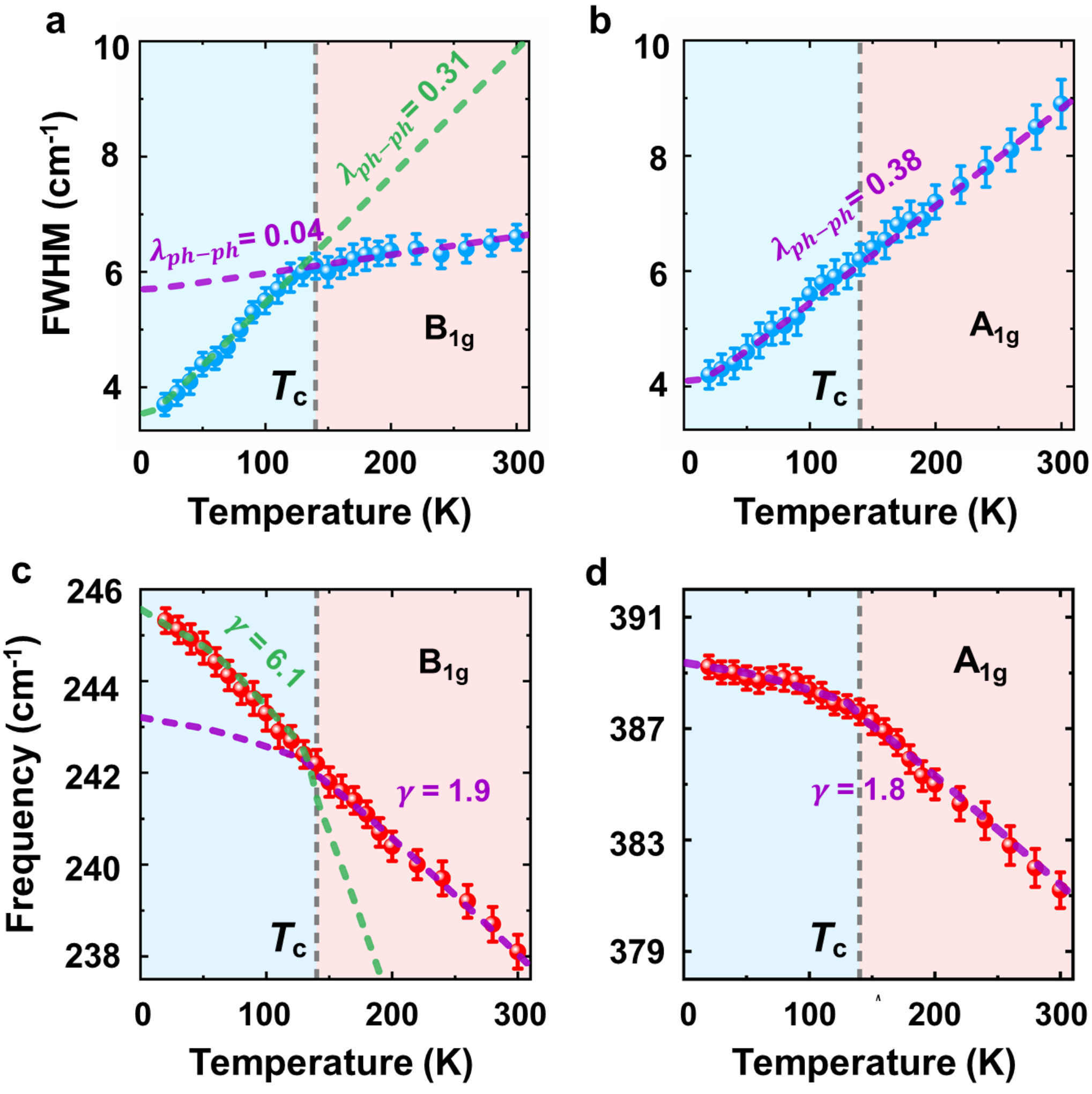

**Figure 4.** Full widths at half maxima (i.e., FWHMs) and energies of the $B_{1g}$ and $A_{1g}$ phonon modes in the Raman spectra of $EuAl_4$ as a function of temperature. **a**) Temperature dependence of the FWHM of the $B_{1g}$ phonon mode. The green and purple dashed lines in **a**) shows the fits to the temperature dependence of the FWHM below and above $T_c$ based on Equation (4). The fitting yields the phonon-phonon interaction (PPI) parameters for the $B_{1g}$ phonon mode: $\lambda_{ph\text{-}ph}$ ~ 0.31 at $T < T_c$ and $\lambda_{ph\text{-}ph}$ ~ 0.04 at $T > T_c$. **b**) Temperature dependence of the FWHM of the $A_{1g}$ phonon mode. The PPI parameter for the $A_{1g}$ phonon mode $\lambda_{ph\text{-}ph}$ ~ 0.38 in the temperature range 15—300 K was obtained via the fitting shown by the purple dashed line in **b**). **c**) Temperature evolution of the $B_{1g}$ phonon energy. The fitting based on Equation (5)-(7) with the PPI parameters for the $B_{1g}$ phonon mode $\lambda_{ph\text{-}ph}$ ~ 0.04 and $\lambda_{ph\text{-}ph}$ ~ 0.31 yields the Grüneisen parameters $\gamma = 1.9$ at $T > T_c$ and $\gamma = 6.1$ at $T < T_c$, respectively. **d**) Temperature evolution of the $A_{1g}$ phonon energy. The green and purple dashed lines in a) shows the fits to the temperature dependence of the $A_{1g}$ phonon energy based on Equation (5)-(7) with $\gamma = 1.8$ and $\lambda_{ph\text{-}ph}$ ~ 0.38.

**Table 1.** Energies of the calculated and measured Raman-active phonon modes $B_{1g}$ and $A_{1g}$ of $EuAl_4$

| Symmetry | Active | Calculated energy [$EuAl_4$] | Experimental energy [$EuAl_4$] |
|---|---|---|---|
| $A_{1g}$ | Raman | 229.7 $cm^{-1}$ | 246 $cm^{-1}$ |
| $B_{1g}$ | Raman | 359.9 $cm^{-1}$ | 390 $cm^{-1}$ |